\documentclass[11pt]{article}
\usepackage{amsmath,amssymb,color,graphics,epsfig}

\makeatletter
\@addtoreset{equation}{section}
\makeatother

\usepackage{amsfonts}
\newcommand{\hoch}[1]{$\, ^{#1}$}

\newcommand{\be}{\begin{equation}}
\newcommand{\ee}{\end{equation}}
\newcommand{\bea}{\setlength\arraycolsep{2pt} \begin{eqnarray}}
\newcommand{\eea}{\end{eqnarray}}
\newcommand{\nn}{\nonumber}

\def\ft#1#2{{\textstyle{\frac{\scriptstyle #1}{\scriptstyle #2} } }}
\def\fft#1#2{{\frac{#1}{#2}}}

\def\0{{\sst{(0)}}}
\def\1{{\sst{(1)}}}
\def\2{{\sst{(2)}}}
\def\3{{\sst{(3)}}}
\def\4{{\sst{(4)}}}
\def\5{{\sst{(5)}}}
\def\6{{\sst{(6)}}}
\def\7{{\sst{(7)}}}
\def\8{{\sst{(8)}}}
\def\sst#1{{\scriptscriptstyle #1}}

\def\ii{{\rm i}}

\begin{document}

\begin{flushright}
%\hfill{KIAS-P12028}
 %\hfill{
%\bf hep-th/yymmnnn}
\end{flushright}

\vspace{25pt}
\begin{center}
{\large {\bf C-metrics in Gauged STU Supergravity and Beyond}}

\vspace{10pt}
H. L\"u\hoch{1} and Justin F. V\'azquez-Poritz\hoch{2}

\vspace{10pt}

{\it \hoch{1}Department of Physics,
Beijing Normal University, Beijing 100875, China}

\vspace{10pt}

\hoch{2}{\it Physics Department, New York City College of Technology\\ The City University of New York, Brooklyn NY 11201, USA}

\vspace{40pt}

\underline{ABSTRACT}
\end{center}

We construct charged generalizations of the dilaton C-metric in various four-dimensional theories, including STU gauged supergravity as well as a one-parameter family of Einstein-Maxwell-dilaton theories whose scalar potential can be expressed in terms of a superpotential. In addition, we present time-dependent generalizations of the dilaton C-metric and dilaton Ernst solutions, for which the time evolution is driven by the dilaton. These C-metric solutions provide holographic descriptions of a strongly-coupled three-dimensional field theory on the background of a black hole, a gravitational soliton, and a black hole undergoing time evolution.

\vfill {\footnotesize Emails: mrhonglu@gmail.com \ \ \ jvazquez-poritz@citytech.cuny.edu}

\thispagestyle{empty}

\pagebreak

\tableofcontents
\addtocontents{toc}{\protect\setcounter{tocdepth}{2}}

%%%%%%%%%%%%%%%%%%%%%%%%%%%%%%%%%%%%%%%%
%\newpage
%%%%%%%%%%%%%%%%%%%%%%%%%%%%%%%%%%%%%%%%

\vspace{4pt}

\section{Introduction}

The C-metric is a well-known vacuum spacetime that describes uniformly accelerating black holes. It was originally discovered in its static form by Hermann Weyl in 1917 \cite{Weyl:1917gp}, although it was not actually coined the C-metric until 1962 \cite{Ehlers:1962zz}. Its analytic extension describes a pair of black holes accelerating away from each other due to the presence of conical singularities, which could be interpreted either as the black holes being pulled by a cosmic string or else being pushed by a strut \cite{Kinnersley:1970zw,bonnor}. For reviews on more recent work pertaining to the C-metric see, for example, \cite{Pravda:2000vh,Letelier:1998rx,c-global}.

 Over the years, various generalizations of the C-metric have been constructed. For example, charged AdS C-metric solutions in Einstein-Maxwell theory with a cosmological constant arise within a subset of the solutions found in \cite{Plebanski:1976gy}. Moreover, dilatonic generalizations of the charged C-metric for an arbitrary dilaton coupling parameter were found in \cite{Dowker:1993bt}. These solutions have recently been generalized in two directions in \cite{Lu:2014ida}. Firstly, charged dilaton C-metrics have been found in Einstein-Maxwell-dilaton theories which have a scalar potential that can be expressed in terms of a superpotential. Secondly, an exact time-dependent charged dilaton C-metric in four-dimensional ${\cal N}=4$ gauged supergravity has been constructed. The dilaton drives the time evolution by transferring its energy to the black holes, which causes their masses to increase and their acceleration to decrease.

C-metrics are connected with a number of developments in string theory and, more generally, gravity in higher dimensions. For instance, a charged dilaton C-metric can be lifted to five dimensions where, upon double Wick rotation, it becomes a rotating black ring \cite{Emparan:2001wn}. The existence of such a five-dimensional black hole with an event horizon of topology $S^1\times S^2$ shows that there is no obvious analog of the uniqueness theorems in five dimensions. AdS C-metrics have been used to construct localized black holes on a UV brane in the brane-world context \cite{Emparan:1999wa,Emparan:1999fd} and plasma ball solutions on an IR brane \cite{Emparan:2009dj}.
Also, the AdS/CFT correspondence \cite{Aharony:1999ti} has been applied to the AdS C-metrics to provide a holographic description of a strongly-coupled three-dimensional conformal field theory on a black hole background \cite{Hubeny:2009ru,Hubeny:2009kz}.

We explore various avenues for further generalizing C-metric solutions. This includes multiply-charged C-metric solutions in the STU gauged supergravity model, as well as two-charge C-metrics in a one-parameter family of Einstein-Maxwell-dilaton theories whose scalar potential can be expressed in terms of a superpotential. We also present exact time-dependent dilaton C-metric and dilaton Ernst solutions in ${\cal N}=4$ gauged supergravity, which describe evolving black holes. In the case of the dynamical dilaton Ernst solution, one can tune the value of the magnetic field such that the cosmic string can be completely removed for all times. The metrics on the boundary of the C-metric solutions describe three-dimensional static black holes, gravitational solitons, as well as black holes undergoing time evolution. According to the AdS/CFT correspondence, the dual three-dimensional conformal field theory lives on these various boundary geometries.

In the next section, we begin by discussing the basic properties of the C-metric solutions in Einstein-Maxwell theory with a cosmological constant. Then we construct charged C-metric solutions in STU gauged supergravity in section 3, and for a one-parameter family of Einstein-Maxwell-dilaton theories in section 4. In section 5, we discuss the dynamical generalizations of the dilaton C-metric and dilaton Ernst solutions. The geometry on the boundary of all of the C-metric solutions is analyzed in section 6. In section 7, we discuss the prospect of higher-dimensional C-metrics, and lift a dilaton C-metric to five dimensions where it is a Ricci-flat spacetime. Lastly, our conclusions are presented in section 8.

\section{C-metrics in Einstein-Maxwell theory}

In this section, we review the C-metric solutions in Einstein-Maxwell theory with a
cosmological constant
\be
e^{-1}{\cal L} = R - \ft14 F^2 - 2\Lambda\,,
\ee
where $F=dA$. Throughout this paper, we will consider a negative cosmological constant parameterized by $\Lambda=-3g^2$, where the constant $g$ is associated with the gauge coupling in gauged supergravities.  The C-metric is a special case of the general Pleba\'nski-Demia\'nski solution
\cite{Plebanski:1976gy}. The ansatz takes the form
\bea
ds^2 &=& \fft{1}{\alpha^2(x-y)^2}\Big[-Y(y) d\tau^2 + \fft{dy^2}{Y(y)} +
\fft{dx^2}{X(x)} + X(x) d\varphi^2\Big]\,,\nn\\[2mm]
A &=&4Q\,y d\tau + 4P\,x d\varphi\,.\label{emC}
\eea
The Maxwell field $A$ carries both electric and magnetic charges parameterized by $Q$ and $P$ respectively. The ansatz is consistent with the equations of motion and the functions $X$ and $Y$ can be solved exactly, given by
\be
X(x)=\Xi(x)\,,\qquad
Y(y)=-\Xi(y) + \fft{g^2}{\alpha^2}\,,
\ee
with
\be
\Xi(\xi) = b_0 + b_1 \xi + b_2 \xi^2 + b_3 \xi^3 -4\alpha^2(Q^2 + P^2) \xi^4
\,.
\ee
The form of the general solution is invariant under the simultaneous constant shift and scaling of the $(x,y)$ coordinates. Together with the overall trombone scaling symmetry of the equations of motion, (i.e. $g_{\mu\nu} \rightarrow \lambda^2 g_{\mu\nu}, A_\mu\rightarrow \lambda A_\mu$, and the cosmological constant is allowed to scale appropriately), one can get rid of three parameters and the solution can be expressed as
\be
\Xi(\xi) = (1-\xi^2) \Big(1 + 2m\alpha \xi + 4\alpha^2(Q^2 + P^2) \xi^2\Big)
\,.
\ee

The physical interpretation of these C-metric solutions was studied in detail in \cite{Dias:2002mi,Podolsky:2003gm,Krtous:2005ej}. To understand the physical meaning of the parameters, it is advantageous to make the coordinate transformation
\be
y=-\fft1{\alpha r}\,,\qquad \tau = -\alpha t\,.\label{ytor}
\ee
Then the solution (\ref{emC}) can be expressed as
\bea
ds^2 &=& \fft{1}{(1 + \alpha x r)^2} \left[-\Delta_r dt^2 + \fft{dr^2}{\Delta_r} + r^2\left(\fft{dx^2}{X} + X d\varphi^2\right)\right]\,,\nn\\[2mm]
A &=&\fft{4Q}{r}\ dt + 4P\,x d\varphi\,,\cr
\Delta_r &=&g^2 r^2 + (1-\alpha^2 r^2) \left(1 - \fft{2m}{r} + \fft{4(Q^2+P^2)}{r^2}\right)\,,\cr
X &=& (1-x^2) \left(1+ 2m\alpha x + 4\alpha^2 (Q^2 + P^2) x^2\right)\,.
\label{emc1}
\eea
The physical interpretation becomes clearer.  For $\alpha=0$, the solution reduces to the well-known Reissner-Nordstr\"om AdS back hole for which $Q$ and $P$ are the electric and magnetic charges.  For vanishing mass and charges, the solution describes AdS spacetime in the accelerating frame, where $\alpha$ is the acceleration parameter.  For $g=0$, the acceleration Killing horizon and the black hole event horizon can be determined by the vanishing of the quantities within the first and second set of parentheses in $\Delta_r$, respectively.  Both horizons are shifted by the introduction of the cosmological constant.  The global structure of the Ricci-flat C-metrics, corresponding to $g=0$ and vanishing charges, was thoroughly analyzed in \cite{c-global}.

The coordinate range for $x$ lies in the region $[x_1,x_2]$ for which $X$ is non-negative and has a maximum, where $x_{1,2}$ are two adjacent simple roots of $X$.  We can always make simultaneous constant shifting and scaling of the coordinates $(x,y)$ so that the region is within any specified interval.  Thus, we choose the parameters appropriately so that the coordinate $x=\cos\theta$ in (\ref{emc1}) can lie in the closed interval $[-1,1]$; it becomes the polar coordinate on a smooth 2-sphere when $\alpha=0$, provided that $\varphi$ has the period $2\pi$.  This can be achieved if the second bracket of $X$ in (\ref{emc1}) has no real roots, provided that
\be
0\le m <2\sqrt{Q^2 + P^2}\,.\label{cond1}
\ee
Alternatively, the second bracket of $X$ can have two real roots that are both less than $-1$ provided that
\be
0\le 2m \alpha < 1+ 4\alpha^2 (Q^2 + P^2)<2\,,\label{cond2}
\ee
which may overlap with condition (\ref{cond1}).  It should be emphasized that these requirements are inessential but sometimes convenient.

There is an additional hidden variable $C$ that specifies the period of $\varphi$, i.e. $\Delta\varphi=2\pi C$.  This is due to the fact that one can generally adjust the parameter $C$ to smooth out only one conical singularity, either at the south or north pole but not both.  The exception to this rule is when the mass parameter $m$ vanishes, in which case the metric becomes an even function of $x$ and hence both the north and south pole can be simultaneously made to be smooth. However, this comes with the price of not having an event horizon, so that the curvature singularity at $y=\infty$ becomes naked.  (Note that the singularity is naked also when the condition (\ref{cond1}) is satisfied.) Interestingly enough, as we shall see in section 6, the boundary metric can be a bounded smooth soliton in this case.

As discussed in \cite{Dias:2002mi,Podolsky:2003gm,Krtous:2005ej}, for $\alpha<g$ the C-metric represents a single uniformly accelerating black hole within an asymptotically AdS spacetime. $\alpha=g$ corresponds to accelerating black holes entering and leaving the asymptotically AdS universe through its conformal infinity one at a time. For $\alpha>g$, the solution describes a sequence of pairs of uniformly accelerating black holes. Interestingly enough, the negative cosmological constant induces an effective attractive force between the two black holes. Thus, the magnitude of the string/strut tension must be larger than a critical value in order to keep the pair of black holes accelerating apart. We find that these properties persist for all of the new C-metric solutions found in this paper.

The solution (\ref{emc1}) carries both electric and magnetic charges, given by
\be
Q_e=\fft{1}{16\pi} \int {*F}=C\,Q\,,\qquad
Q_m=\fft{1}{16\pi} \int F=C\,P\,.
\ee
However, we will not always be pedantic about the $C$ factor in referring to the charges.

\section{C-metrics in STU gauged supergravity}

The STU supergravity model can be obtained from $D=6$ string theory.  It has $SL(2,R)\times SL(2,R)\times SL(2,R)$ global symmetry, corresponding to-- when discretized-- the (S,T,U) duality symmetries of the string theory \cite{dulura}. The theory can also be gauged to become STU gauged supergravity with ${\cal N}=2$ supersymmetry (further details of the theory can be found in \cite{Duff:1999gh,Cvetic:1999xp,Lu:2014fpa}). The theory contains four $U(1)$ vector fields $F_I=dA_I$ and three complex scalars $\tau_i=\chi_i + {\rm i} e^{\phi_i}$.   The axions $\chi_i$ can be further truncated out, provided that
\be
F_I\wedge F_J=0\,,\qquad I\ne J\,.\label{fijcons}
\ee
The truncated Lagrangian is given by
\bea
e^{-1} {\cal L} =R -\ft12 \sum_{i=1}^3 (\partial \phi_i)^2
 - \ft14 \sum_{I=1}^4 e^{\vec a_I\cdot \vec \phi} F_I^2+
2g^2 \sum_{i=1}^3 \cosh\phi_i\,.
\eea
where
\be
\vec a_1 = (1,1,1)\,,\qquad \vec a_2=(1,-1,-1)\,,\qquad
\vec a_3=(-1,1,-1)\,,\qquad \vec a_4=(-1,-1,1)\,.
\ee
This theory can be obtained from eleven-dimensional supergravity by a consistent reduction on a seven-sphere \cite{Cvetic:1999xp}.  The $U(1)^4$-charged black holes were constructed in \cite{Duff:1999gh,Sabra:1999ux}.  They can be lifted to eleven dimensions where they become the near-horizon geometry of the rotating M2-brane \cite{Cvetic:1999xp}.

To construct the charged C-metrics in the STU gauged supergravity model, we consider the ansatz
\bea
ds^2 &=& \fft{1}{\alpha^2(y-x)^2} \left[ \sqrt{{\cal F}} \left(-Y d\tau^2 + \fft{dy^2}{Y}\right) + \sqrt{{\cal H}} \left(\fft{dx^2}{X} + X d\varphi^2\right)\right]\,,\nn\\[2mm]
e^{\vec a_I \cdot \vec\phi} &=& U_I^{2} (U_1U_2U_3U_4)^{-\fft12}\,,
\qquad F_I=\fft{4Q_I}{h_I^2} dy\wedge d\tau\,,\qquad
U_I=\fft{h_I}{f_I}\cr
f_I &=&1 + \alpha q_I x\,,\qquad h_I=1 + \alpha q_I y\,,\qquad {\cal F}\equiv \prod_{I=1}^4f_I\,,\qquad {\cal H}\equiv \prod_{I=1}^4h_I\,,
\eea
where the functions $X$ and $Y$ depend on the coordinates $x$ and $y$, respectively.  The ansatz is motivated by the structure of the $U(1)^4$-charged black holes, as well as the general structure of C-metrics discussed in section 2 and refs.~\cite{Dowker:1993bt,Lu:2014ida}.

The Bianchi identities for the field strengths are clearly satisfied.  To check the Maxwell equations, we note that
\be
e^{\vec a_I \cdot \vec \phi} {*F_I} = \fft{4Q_I}{f_I^2} dx\wedge d\varphi\,,
\label{stustarF}
\ee
which all vanish under the exterior derivative, since the $f_I$ are functions of $x$ only.  The electric charges are given by
\be
Q^e_I = \fft{Q_I}{4\pi} \int \fft{dx}{f_I^2} \int d\varphi\,.\label{stuQe}
\ee
Thus, in addition to the $Q_I$ and the hidden parameter $C$ that specifies the period of $\varphi$, the conserved electric charges also depend on the scalar charge parameters $q_I$.  This is in contrast with the C-metric of Einstein-Maxwell theory discussed in the previous section.  It is worth pointing out that it follows (\ref{stustarF}) that we can also consider the magnetically-charged ansatz
\be
F_I = \fft{4Q_I}{f_I^2} dx\wedge d\varphi\,.
\ee
It is clear that the equations of motion are identical provided that the dilatons are now given by
\be
e^{\vec a_I \cdot \vec\phi} = U_I^{-2} (U_1U_2U_3U_4)^{\fft12}\,.
\ee
This Hodge dualization can be done to any number of the field strengths, whilst the metric remains the same.

The three scalar equations turn out to be enough to determine the $X$ and $Y$ functions completely, with an additional integration constant $b_0$.  It is then a straightforward exercise to verify the Einstein equations.  We find that
\be
X= \fft{\widetilde X}{\sqrt{{\cal F}}}\,,\qquad
Y=-\fft{\widetilde Y}{\sqrt{{\cal H}}} + \fft{g^2}{\alpha^2} \sqrt{{\cal H}}\,,
\ee
where
\bea
\widetilde X &=& {\cal F}\left( b_0 + \sum_{I=1}^4 \fft{16Q_I^2}{\alpha^2 f_Iq_I \prod_{J\neq I} (q_J-q_I)}\right)\,,\nn\\[2mm]
\widetilde Y &=& {\cal H} \left( b_0 + \sum_{I=1}^4 \fft{16Q_I^2}{\alpha^2 h_Iq_I \prod_{J\neq I} (q_J-q_I)}\right)\,.
\label{stuxygen}
\eea
The general solution contains the 10 parameters: $(\alpha, b_0, q_I,Q_I)$ where $I=1,\dots , 4$.  The parameter $b_0$ can be scaled to have the discrete values $(-1,0,1)$.  One of the charge parameters $q_I$ can be set to zero by a coordinate transformation of $(x,y)$, on which we will elaborate in the next section (the STU model has three independent scalars, after all).  In our construction, the parameters $Q_I$ are independent from the scalar charges $q_I$.  Therefore, we could set $Q_I=0$ and obtain C-metric solutions supported solely by the scalar charges. On the other hand, for generic $Q_I$ we cannot set the scalar charges to zero, as one would have expected from the theory.  When all of the electric charge parameters are equal, namely $Q_I\equiv Q$, we can then set all $q_I\equiv q$, the three scalar fields all vanish and we recover the charged C-metric solution of the Einstein-Maxwell theory. 

For the C-metric solution in Einstein-Maxwell theory and the dilaton C-metric in \cite{Dowker:1993bt}, conical singularities can be avoided only for a single point in parameter space. In contrast, conical singularities can be completely absent for a range of parameters for the charged C-metric solutions in STU gauged supergravity.

In order to make contact with the charged black holes in the literature, we relate the electric and
scalar charges as follows:
\be
q_I=-\mu s_I^2\,,\qquad Q_I=\ft14 \mu s_I c_I\,,\qquad
s_I=\sinh\delta_I\,,\qquad c_I=\cosh\delta_I\,.
\ee
We find that $\widetilde X$ and $\widetilde Y$ simplify dramatically and become
\be
\widetilde X=b_0 {\cal F} - x^2 (1 + \alpha \mu x)\,,\qquad
\widetilde Y=b_0 {\cal H} - y^2 (1 + \alpha \mu y)\,.
\ee
Making the coordinate transformation (\ref{ytor}) and setting $b_0=1$, the charged C-metric solutions become
\bea
ds^2 &=& \fft{1}{(1 + \alpha r x)^2} \left[ \sqrt{{\cal F}} \left(-\Delta_r dt^2 + \fft{dr^2}{\Delta_r}\right) + \sqrt{{\cal H}} r^2
\left(\fft{dx^2}{X} + X d\varphi^2\right)\right]\,,\nn\\[2mm]
e^{\vec a_I \cdot \vec\phi} &=& U_I^{2} (U_1U_2U_3U_4)^{-\fft12}\,,
\qquad A_I=\fft{\mu\, s_I c_I}{r h_I} dt\,,\qquad
U_I=\fft{h_I}{f_I}\,,\nn\\[2mm]
X &=& {\cal F}^{-\fft12}\Big( {\cal F} - x^2 (1 + \alpha\mu x)\Big)\,,\qquad
\Delta_r = {\cal H}^{-\fft12} \Big(1 - \fft{\mu}{r} + (g^2-\alpha^2)r^2 {\cal H}\Big)\,,\nn\\[1mm]
f_I &=& 1 -\alpha\mu s_I^2 x\,,\qquad h_I = 1 + \fft{\mu s_I^2}{r}\,.\label{stuxr}
\eea
Setting the acceleration parameter to zero precisely gives rise to the charged AdS black holes obtained in \cite{Duff:1999gh,Sabra:1999ux}.

   It is worth mentioning that maximal supergravity in four dimensions is not unique; there exists a one-parameter family extension, called $\omega$-deformed $SO(8)$ gauged supergravity \cite{Dall'Agata:2012bb,deWit:2013ija}. The $\omega$-deformed STU gauged supergravity was given in
\cite{Lu:2014fpa}.  Following the procedure outlined in \cite{Lu:2014fpa}, the charged C-metrics in $\omega$-deformed STU gauged supergravity can be easily obtained.

The solution (\ref{stuxr}) indicates that one needs to be careful as to how to specialize the general form of the solution (\ref{stuxygen}) without loss of generality.  In particular, if we simply set $Q_I=0$, we shall not recover the neutral C-metric, whilst we can for (\ref{stuxr}) by setting $s_I=0$.  Similarly, if we set pair-wise charges equal, or three charges equal, then we obtain a reduced system involving only one scalar.  The result appears to be less general than the previously-known results.  For this reason, we shall consider the single-scalar system on its own merit. We shall carry this out in the next section, which allows us to obtain further charged C-metrics in theories that are inspired by but go beyond supergravity theories.

\section{Further charged C-metrics}

The STU supergravity model discussed in the previous section can be reduced to involve only one scalar when the four Maxwell vectors are set pair-wise equal, or else when three vectors are set equal. It turns out that these theories can be generalized to a one-parameter family of theories given by
\be
e^{-1}{\cal L} = R - \ft12 (\partial\phi)^2 - \ft14 e^{a_1\phi} F_1^2 -
\ft14 e^{a_2\phi} F_2^2 - V(\phi)\,,\label{lag2}
\ee
where the dilaton coupling constants $a_1$ and $a_2$ are
\be
a_1=\sqrt{\ft{N_2}{N_1}}\,,\qquad a_2=-\sqrt{\ft{N_1}{N_2}}\,,
\qquad \hbox{with}\qquad N_1 + N_2=4\,.
\ee
The scalar potential can be expressed in terms of a superpotential \cite{Lu:2013eoa}:
\be
V = \left(\fft{dW}{d\phi}\right)^2-\ft34 W^2\,,\qquad
W = \ft{1}{\sqrt2} g\left( N_1 e^{-\fft12 a_1\phi}+ N_2 e^{-\fft12 a_2\phi}\right)\,.
\ee
The scalar potential has a stationary point at $\phi=0$ with $V_0=-6g^2$, a negative cosmological constant.  The theory with positive integers $N_i$ can arise from the STU gauged supergravity via a consistent truncation.  The higher-dimensional generalization of this theory along with the corresponding charged AdS black holes were given in \cite{Lu:2013eoa}.

   In this section, we shall construct the two-charge C-metric solutions of the theory (\ref{lag2}). Inspired by the AdS black hole solutions constructed in \cite{Lu:2013eoa} along with the general structure of C-metrics, we consider the ansatz
\bea
ds^2 &=& \fft{1}{\alpha^2(y-x)^2} \left[ f_1^{\fft12 N_1} f_2^{\fft12 N_2} \left(-Y d\tau^2 +
\fft{dy^2}{Y}\right) + h_1^{\fft12 N_1} h_2^{\fft12 N_2} \left(\fft{dx^2}{X} + X d\varphi^2\right)\right]\,,\nn\\[2mm]
e^{\phi} &=& \left(\fft{h_1\,f_2}{f_1\,h_2}\right)^{\fft12\sqrt{N_1N_2}}\,,
\qquad F_1=\ft{4Q_1}{h_1^2}\ dy\wedge d\tau\,,\qquad
F_2=\ft{4Q_2}{h_2^2}\ dy\wedge d\tau\,,\nn\\[2mm]
f_i &=&1 + \alpha q_i x\,,\qquad h_i=1 + \alpha q_i y\,.\label{twochargeans}
\eea
This ansatz involves two electric charge parameters $(Q_1,Q_2)$ and two scalar charges $(q_1,q_2)$. (As in the case of STU gauged supergravity, the magnetically-charged solutions can be easily obtained.) Since we only have one scalar field, we expect that one of the scalar charges can be removed through an appropriate coordinate transformation. Indeed, the ansatz is invariant under the transformation
\be
x= \fft{\tilde x}{1 + b \tilde x}\,,\qquad y = \fft{\tilde y}{1 + b \tilde y}\,,\label{xytrans}
\ee
%To see this in some detail, we give
%\bea
%&&\fft{1}{(x-y)^2} = \fft{(1+ b\tilde x)^2 (1 + b \tilde y)^2}{(\tilde x-
%\tilde y)^2}\,,\qquad dx^2 = \fft{d\tilde x^2}{(1 + b \tilde x)^4}\,,\quad
%dy^2 = \fft{d\tilde y^2}{(1 + b \tilde y)^4}\,,\cr
%&&f_i = \fft{1 + (b+\alpha q_i) \tilde x}{1 + b \tilde x}\equiv
%\fft{\tilde f_i}{1 + b\tilde x}\,,\qquad
%h_i = \fft{1 + (b+\alpha q_i) \tilde y}{1 + b \tilde y}\equiv
%\fft{\tilde h_i}{1 + b\tilde y}\,,\cr
%&&f_1^{\fft12 N_2} f_2^{\fft12 N_1} = \fft{\tilde f_1^{\fft12 N_2}
%\tilde f_2^{\fft12 N_1}}{(1+ b \tilde x)^2}\,,\qquad
%h_1^{\fft12 N_2} h_2^{\fft12 N_1} = \fft{\tilde h_1^{\fft12 N_2}
%\tilde h_2^{\fft12 N_1}}{(1+ b \tilde y)^2}\,,\cr
%&&X = \fft{\widetilde X}{(1 +b \tilde x)^2}\,,\qquad Y=
%\fft{\widetilde Y}{(1 + b \tilde y)^2}\,.
%\eea
allowing us to set one of the $q_i$ to zero for an appropriate value of $b$. This coordinate transformation also enables us to set one of the four $q_I$ to zero in the STU C-metric solutions obtained in the previous section. For later purposes as well as for maintaining a symmetric form, we keep both $q_i$ in our ansatz.  This is the same strategy that was used for constructing scalar-hairy black holes in \cite{Feng:2013tza}.

The functions $X$ and $Y$ can be fully solved and we find
\bea
X&=&\left[(b_0 + b_1 x + b_2 x^2) f_1 f_2 + \ft{16}{\alpha^2(q_2-q_1)}\left(
\ft{Q_1^2}{N_1q_1^3} f_2 - \ft{Q_2^2}{N_2q_2^3} f_1\right)\right]
f_1^{-\fft12N_1} f_2^{-\fft12N_2}\,,\nn\\[2mm]
Y&=&-\left[(b_0 + b_1 y + b_2 y^2) h_1 h_2 + \ft{16}{\alpha^2(q_2-q_1)}\left(
\ft{Q_1^2}{N_1q_1^3} h_2 - \ft{Q_2^2}{N_2q_2^3} h_1\right)\right]
h_1^{-\fft12N_1} h_2^{-\fft12N_2}\nn\\[1mm]
&&\qquad
+\fft{g^2}{\alpha^2}\, h_1^{\fft12 N_1} h_2^{\fft12 N_2}\,.
\eea
The full solution contains the eight parameters $(\alpha, b_0,b_1,b_2,q_1,q_2,Q_1,Q_2)$.  The conserved electric charges are
given by
\bea
Q_1^e &=& \fft{Q_1}{4\ii} \int e^{a_1\phi} {*F_1} =
\fft{Q_1}{4\pi} \int \fft{dx}{f_1^2} \int d\varphi\,,\nn\\[2mm]
Q_2^e &=& \fft{Q_2}{4\pi} \int e^{a_2\phi} {*F_2} =
\fft{Q_2}{4\pi} \int \fft{dx}{f_2^2} \int d\varphi\,.\label{2chargeQe}
\eea

  As previously discussed, one of the two $q_i$ can be set to zero. Two more parameters can be removed by way of a constant shift and scaling of the $(x,y)$ coordinates. Moreover, another parameter can be fixed via the overall trombone scaling symmetry, provided that the gauge constant $g$ is allowed to be scaled.  Thus, without loss of generality, we can write the function $X$ as
\be
X=(1-x^2) (1 + \alpha p_1 x + \alpha^2 p_2 x^2)f_1^{-\fft12N_1} f_2^{-\fft12N_2}\,.\label{specifyX}
\ee
This yields five algebraic equations and, since we have five parameters $(b_0,b_1,b_2,p_1,p_2)$, a solution is guaranteed.  We find that
\bea
p_1 &=& q_1 + q_2 + \ft{1}{q_2-q_1}\Big(\ft{16 Q_1^2}{N_1(1-\alpha^2 q_1^2)}
-\ft{16 Q_2^2}{N_2(1-\alpha^2 q_2^2)}\Big)\,,\nn\\[2mm]
p_2 &=& q_1 q_2 + \ft{1}{q_2-q_1}\Big(\ft{16 q_2 Q_1^2}{N_1 (1-\alpha^2 q_1^2)} - \ft{16 q_1 Q_2^2}{N_2(1-\alpha^2 q_2^2)}\Big)\,.
\eea
It is then straightforward to verify that $Y$ can be expressed as
\be
Y=-(1-y^2) (1 + \alpha p_1 y + \alpha^2 p_2 y^2)h_1^{-\fft12N_1} h_2^{-\fft12N_2}+\fft{g^2}{\alpha^2}\, h_1^{\fft12 N_1} h_2^{\fft12 N_2}\,.
\ee
Making the coordinate transformation (\ref{ytor}), the C-metric solution can be written as
\bea
ds^2 &=& \fft{1}{(1+ \alpha r x)^2} \left[ f_1^{\fft12 N_1} f_2^{\fft12 N_2} \left(-\Delta_r dt^2 +
\fft{dr^2}{\Delta_r}\right) + h_1^{\fft12 N_1} h_2^{\fft12 N_2} r^2\left(\fft{dx^2}{X} + X d\varphi^2\right)\right]\,,\nn\\[2mm]
e^{\phi} &=& \left(\ft{h_1\,f_2}{f_1\,h_2}\right)^{\fft12\sqrt{N_1N_2}}\,,
\qquad F_1=\ft{4Q_1}{r^2 h_1^2}\ dr\wedge dt\,,\qquad
F_2=\ft{4Q_2}{r^2 h_2^2}\ dr\wedge d t\,,\nn\\[2mm]
f_i &=&1 + \alpha q_i x\,,\qquad h_i=1 - \fft{q_i}{r}\,,\nn\\[2mm]
X&=&(1-x^2) (1 + \alpha p_1 x + \alpha^2 p_2 x^2)f_1^{-\fft12N_1} f_2^{-\fft12N_2}\,,\nn\\[2mm]
\Delta_r &=& (1-\alpha^2 r^2)\left(1- \fft{p_1}{r} + \fft{p_2}{r^2}\right)
h_1^{-\fft12N_1} h_2^{-\fft12N_2}+g^2 r^2\, h_1^{\fft12 N_1} h_2^{\fft12 N_2}\,.\label{solution2}
\eea
For vanishing $\alpha$, this reduces to the charged AdS black hole solutions \cite{Lu:2013eoa}.
Without loss of generality, we can set $q_1=q$ and $q_2=0$. Then for vanishing $Q_1$, $X$ can be factorized with a factor $f$ and the solution reduces to the one that was obtained in \cite{Lu:2014ida}.  Note that the solution with vanishing $Q_2$ appears to be more complicated but is actually equivalent to that with $Q_1=0$ by a coordinate transformation of the type (\ref{xytrans}).

It is worth remarking that, in the above parametrization, the scalar and the electric charges are independent whilst the mass parameter is not manifest.  We can set $Q_i=0$ and obtain the C-metrics for the pure scalar sector.  An alternative and standard black hole parametrization is given by
\be
Q_i=\ft14 \mu \sqrt{N_i} s_i c_i\,,\qquad
q_i=-\mu s_i^2\,,\qquad
s_i=\sinh\delta_i\,,\quad c_i=\cosh\delta_i\,.\label{musici}
\ee
In this parametrization, it is easy to verify that for $\alpha=0$ we have
\be
p_1=-\mu\,,\qquad p_2=0\,.
\ee
Thus, $\mu$ becomes the mass parameter whilst $Q_i$ and $q_i$ are mixed together and, consequently, it becomes less obvious as to how to truncate the solution to the pure scalar system. Nevertheless, this does have the advantage that one can obtain the Schwarzschild black hole simply by setting $\delta_i=0$.  In fact, the supergravity solutions can be obtained through a solution-generating technique in which the $\delta_i$ are related to Lorentz boosts.  However, this parametrization is not useful for the purpose of generalizing the solutions to become dynamical.  This is because, in a dynamical solution, the electric charges and scalar charges must be distinguished.  The former ones are conserved whilst the latter ones are not and can evolve in time.  In fact, the scalar field supplies the driving force for the dynamical geometry.

\section{Dynamical solutions in ${\cal N}=4$ gauged supergravity}

\subsection{Dynamical C-metric}

In this section, we construct dynamical charged C-metrics.  Our starting point is the one-scalar system (\ref{lag2}), whose static solution takes the form (\ref{twochargeans}).  It is easy to see that the $(\tau,y)$ coordinates can be combined to form Eddington-Finkelstein-like coordinates
\be
-Y d\tau^2 + \fft{dy^2}{Y}= -2\eta du dy - Y du^2\,,
\ee
by taking $du = d\tau -\eta Y^{-1}dy$.  We can then promote the metric ansatz to be dependent on the advanced $(\eta=1)$ or retarded $(\eta=-1)$ time coordinate $u$.  We shall preserve the conservation of electric charge whilst the time evolution dynamics are caused by the non-conserved scalar charges. However, this immediately leads to a restriction for incorporating time dependence. It follows from (\ref{2chargeQe}) that the conserved $Q^e_i$ depend not only on $Q_i$ but also on $q_i$.  With our current ansatz, this renders it impossible for both $Q^e_i$ to remain conserved whilst promoting any of the scalar charges $q_i$'s to be $u$-dependent\footnote{It follows from (\ref{stustarF}) and (\ref{stuQe}) that the same obstacle exists for constructing the dynamical C-metrics in the STU model.}.  Letting the parameters $Q_i$ be $u$-dependent, which is permitted by the Bianchi identities, does not help the situation. One way to resolve this is to set one of the electric charges to zero, say $Q_1=0$, and then set $q_2$ to be a constant, which can be made to be zero by a coordinate transformation. This leads to a system with Einstein gravity, a dilaton and a single vector field.

The spherically-symmetric Robinson-Trautman ansatz for collapsing solutions in the
Einstein-Maxwell-dilaton theory were analysed in \cite{Gueven:1996zm}. It was shown that the dilaton coupling constant $a$ must be unity, thereby leading to a solution that can be embedded in ${\cal N}=4$ supergravity.  The collapsing two-charge black hole solution in ${\cal N}=4$ gauged supergravity was obtained in \cite{Lu:2014eta,Zhang:2014sta}.  Since the C-metrics naturally include the black hole solution, we shall consider collapsing (single-charge) C-metrics in ${\cal N}=4$ gauged supergravity.

Taking all of this into account, we consider the Langrangian
\be
e^{-1}{\cal L} = R - \ft12 (\partial\phi)^2  -\ft14 e^{-\phi} F^2 +2g^2 (\cosh\phi + 2)\,.\label{lag3}
\ee
We take the time-dependent ansatz
\bea\label{time-dependent-solution}
ds^2 &=& \fft{1}{\alpha (u)^2(y-x)^2} \left[ f \left( -2\eta du dy-Y(u,x,y) du^2\right) + h \left(\fft{dx^2}{X} + X d\varphi^2\right)\right]\,,\nn\\[2mm]
e^{\phi} &=& \fft{h}{f}\,,\qquad F=4Q\ dy\wedge du\,,\\[1mm]
X &=& (1-x^2) (1 + 2m(u)\alpha(u) x)\,,\qquad f=1 + \alpha(u) q(u)\, x\,,\qquad h=1 + \alpha(u) q(u)\, y\,,\nn
\eea
where the effective mass $m$, acceleration $\alpha$ and scalar charge $q$ depend on $u$ while the electric charge $Q$ remains conserved. The function $Y$ is to be determined by the equations of motion.

We find that there are three branches of solutions, which have been presented and whose properties have been studied in \cite{Lu:2014ida} with the coordinate $r=-1/(\alpha y)$. The first branch has $\alpha(u)=0$, a limit which can be taken using the $r$ coordinate. Then one recovers the single-charge version of exact black hole formation obtained in \cite{Lu:2014eta}.  For non-vanishing $\alpha$, we find that
\be
q(u)=q_0^2 \alpha(u)\,,\qquad m(u)=\fft{4Q^2}{q(u)}\,,
\ee
where $q_0$ is a constant and the function $Y$ is given by
\be
Y = -(1-y^2) (1 +2m\alpha y)+\fft{g^2}{\alpha^2}\, h+\fft{\eta\dot\alpha}{f^2\alpha}\ (y-x)(3f-h)\,.
\ee
Note that the function $X$ remains independent of $u$. Constant $\alpha$ then gives rise to the static solution in the previous section with $Q_1=0$ and $N_1=2$. On the other hand, a $u$-dependent $\alpha$ must satisfy
\be
\dot \alpha = \fft{\eta q_0^2}{\alpha^3}\, (\alpha^4 - \alpha_1^4)(\alpha^2-\alpha_2^2)\,,\label{dotalpha}
\ee
where a dot denotes a derivative with respect to $u$ and
\be
\alpha_1 =\fft{1}{q_0}\,,\qquad \alpha_2=\fft{2\sqrt2\,Q}{q_0^2}\,.
\ee
The two fixed points $\alpha_1$ and $\alpha_2$ correspond to the masses\footnote{Note that $\alpha_1$ and $\alpha_2$, along with $m_1$ and $m_2$, are switched relative to \cite{Lu:2014ida}. This is related to the fact that here we are using the coordinate $y$ instead of $r$.}
\be\label{m1-m2}
m_1=\fft{4Q^2}{q_0}\,,\qquad m_2=\sqrt2\,Q\,,
\ee
where $\alpha_1\le \alpha_2$, $m_1\ge m_2$, and $m_2$ satisfies the BPS condition \cite{Lu:2014ida}. These inequalities imply that $x\in [-m_2^2/m_1^2,1]$ in order for $X$ to be non-negative. The two fixed points lead to three regions of evolution, the details of which have already been discussed in \cite{Lu:2014ida}. The bottom line is that, by appropriately choosing the sign of $\eta$ within each region of time evolution, energy is transferred from the scalar field to the black holes, thereby increasing their masses and lessening their acceleration. Interestingly enough, there are solutions with an initial acceleration parameter $\alpha>g$ and a final acceleration parameter $\alpha<g$. This describes a pair of accelerating black holes that evolves to a single accelerating black hole.

The solution (\ref{time-dependent-solution}) can be lifted to eleven dimensions using the reduction ansatz obtained in \cite{Cvetic:1999au} (see also \cite{Cvetic:1999xp}). The eleven-dimensional metric is given by
\bea
ds_{11}^2 &=& \Delta^{\fft23}\Bigg[ ds_4^2+\fft{4}{g^2}\Bigg( d\xi^2+\fft{1}{f\cos^2\xi+h\sin^2\xi}
\times \Big( h\cos^2\xi\ d\Omega_3^2+f\sin^2\xi\ d\tilde\Omega_3^2\Big)\Bigg)\Bigg]\,,\nn\\[2mm]
d\Omega_3^2 &=& \big( d\psi+\cos\theta d\phi+\ft{1}{\sqrt{2}}gA_\1\big)^2+d\theta^2+\sin^2\theta d\phi^2\,,\nn\\[1mm]
d\tilde\Omega_3^2 &=& (d\tilde\psi+\cos\tilde\theta d\tilde\phi)^2+d\tilde\theta^2+\sin^2\tilde\theta d\tilde\phi^2\,,
\eea
where the conformal factor is
\be
\Delta=\fft{f\cos^2\xi+h\sin^2\xi}{\sqrt{fh}}\,.
\ee
Alternatively, for vanishing $g$ and $Q$, (\ref{time-dependent-solution}) reduces to a solution of Einstein gravity with a free scalar field, which can be lifted to a five-dimensional Ricci-flat spacetime with the metric
\be
ds_5^2 = \fft{1}{\alpha^2(y-x)^2}
\left(\fft{h}{f}\right)^{\fft{1}{\sqrt{3}}}
\left[ f \left( -2\eta du dy-Ydu^2\right) + h \left(\fft{dx^2}{X} + X d\varphi^2\right)\right]+\left(\fft{h}{f}\right)^{-\fft{2}{\sqrt{3}}} dz^2\,.
\ee
Then for the time-dependent branch of solutions, $\alpha$ is given by
\be
\alpha^2=\fft{1}{q_0} \tanh (2u)\,.
\ee

\subsection{Dynamical Ernst solution}

For $g=0$, one can apply an Ehlers-Harrison type transformation \cite{ehlers,harrison} given by equation (3.1) with $a=1$ in \cite{Dowker:1993bt} on the time-dependent C-metric (\ref{time-dependent-solution}) to obtain the time-dependent dilaton Ernst solution
\bea\label{time-dependent-Ernst}
ds^2 &=& \fft{1}{\alpha^2(y-x)^2} \left[ \Lambda f \left( -2\eta du dy-Y du^2\right) + h \left(\fft{\Lambda}{X} dx^2 + \fft{X}{\Lambda} d\varphi^2\right)\right]\,,\nn\\[2mm]
e^{-\phi} &=& \fft{\Lambda h}{f}\,,\qquad A_\1=-\fft{1}{B\Lambda} (1+4BQx)\ d\varphi\,,\nn\\[1mm]
\Lambda &=& (1+4BQx)^2+\fft{2B^2 Xf}{\alpha^2 (y-x)^2}\,,
\eea
where all functions not specified here are the same as for the time-dependent C-metric solution with $g=0$. The parameter $B$ specifies the strength of the magnetic field.

For the static Ernst solutions, it has been shown that the nodal singularities can be removed for an appropriate value of the $B$ parameter \cite{ernst,Dowker:1993bt}. In other words, the magnetic field plays the analogous role as the cosmic string in the C-metric in providing the force necessary to accelerate the black holes. We find that the nodal singularities can be removed for the dynamical Ernst solution (\ref{time-dependent-Ernst}) as well, provided that $B$ and the period of the $\varphi$ coordinate have either one of the following two values:
\bea
B &=& \fft{1}{2\sqrt{2}\ m_2} \left( \fft{\sqrt{2}\ m_1^2\mp m_1\sqrt{m_1^2-m_2^2}}{\sqrt{2}\ m_2^2\pm m_1\sqrt{m_1^2-m_2^2}}\right)\,,\nn\\[2mm]
\Delta\varphi &=& \fft{4\pi m_2^2(m_1^2+m_2^2)}{\left( \sqrt{2}\ m_2^2\pm m_1\sqrt{m_1^2-m_2^2}\right)^2}\,,
\eea
where the masses $m_1$ and $m_2$ corresponding to the two fixed points are given by (\ref{m1-m2}). It is a nice feature that these values of $B$ and $\Delta\varphi$ are independent of time, so that the cosmic string is absent for all times.

\section{Black holes on the boundary}

For all of the AdS C-metric solutions constructed in this paper, the asymptotic region is located at $y=x$.  The boundary metrics describe a spatially compact universe with Killing horizons. As in the case of AdS C-metrics in pure Einstein gravity considered in \cite{Hubeny:2009ru,Hubeny:2009kz}, according to the AdS/CFT correspondence, these solutions describe a strongly-coupled three-dimensional conformal field theory on black hole backgrounds.

\subsection{Einstein-Maxwell theory}

For non-zero $g$, the boundary metric of the $C$-metric solutions in (\ref{emC}) are given by
\bea
ds_3^2 &=& \fft{dx^2}{\widetilde X X} - \fft{g^2}{\alpha^2} \widetilde X d\tau^2 + X d\varphi^2\,,\cr
X &=& (1-x^2)\Big(1 + 2m\alpha\, x + 4\alpha^2 (Q^2 + P^2) x^2\Big)\,,\qquad
\widetilde X = 1 - \fft{\alpha^2}{g^2} X\,,
\eea
where we have removed a conformal factor and the parameters satisfy the conditions (\ref{cond1}) and/or (\ref{cond2}).  Thus, the three-dimensional space is bounded with no asymptotic infinity.  Let $X^{\rm max}$ denote the maximum value of $X$ in the region of $-1\le x\le 1$. For $g^2/\alpha^2>X^{\rm max}$, the metric component $g_{\tau\tau}$ is finite and positive definite in $x\in [-1,1]$. The metric then has an unavoidable conical singularity at either the north pole at $x=1$ or the south pole at $x=-1$, except when $m=0$. It is rather intriguing that, for vanishing $m$, the bulk spacetime has a naked power-law singularity while the boundary metric can be that of a smooth soliton.

For $g^2/\alpha^2\le X^{\rm max}$, there is a Killing horizon at $x=x_0$ with $x_0\in (-1,1)$ for which $\widetilde X$ vanishes. We can then choose the period of $\varphi$ appropriately so that the metric is smooth at the north pole, and the conical singularity at the south pole is hidden by the horizon at $x=x_0$. We no longer need to impose special conditions in the region inside the horizon $x<x_0$. Therefore, we can relax the conditions (\ref{cond1}) and/or (\ref{cond2}) since we now need to ensure that there is no conical singularity only in the region on or outside the horizon $x_0\le x\le 1$, rather than within the full $[-1,1]$ region.

There can also exist the situation in which $x_0$ is a double root of $\widetilde X$, corresponding to an extremal black hole.  In this case, the temperature vanishes and the near-horizon geometry is AdS$_2\times S^1$.  This can be achieved if the mass and charge parameters are given by
\be
m =\fft{1}{\alpha x_0} + \fft{g^2 (2x_0^2-1)}{\alpha^3 x_0 (x_0^2-1)^2}\,,\qquad
Q^2+P^2 = \fft{1}{4\alpha^2 x_0^2} + \fft{g^2(3x_0^2-1)}{4\alpha^4 x_0^2(x_0^2-1)^2}\,.
\ee
Then the function $\widetilde X$ becomes
\be
\widetilde X = \fft{(x-x_0)^2}{x_0^2} \left( \fft{\alpha^2}{g^2} (x^2-1) +
\fft{(3x_0^2-1) x^2 +2x x_0^3 + (x_0^2-1)^2}{(x_0^2-1)^2}\right)\,.
\ee
For there to be a genuine event horizon at $x=x_0$, one must ensure that there are no additional roots of either $X$ or $\widetilde X$ within the region $(x_0,1)$.

Studying the thermal competition of the black funnels/droplets described by these C-metric solutions provides insight into the properties of the dual field theory at strong coupling. The neutral AdS C-metric describes configurations that are not in thermal equilibrium \cite{Hubeny:2009ru,Hubeny:2009kz}. On the other hand, the charge parameter of the charged AdS C-metric can be tuned such that thermal equilibrium is achieved, for which the planar and droplet horizons of the black droplet configuration are associated with equal temperature \cite{Caldarelli:2011wa}. However, neither the charged AdS C-metric nor any of the other C-metric solutions describe the situation of full equilibrium, for which the horizons are associated with equal temperature as well as equal chemical potential.

\subsection{STU gauged supergravity}

Up to a conformal factor, the three-dimensional metric on the boundary at $y=x$ is
\be
ds_3^2 = \fft{dx^2}{\widetilde X X\cal F} - \fft{g^2}{\alpha^2}\widetilde X d\tau^2 + X d\varphi^2\,,
\ee
where
\be
X =  b_0 + \sum_{I=1}^4 \fft{16Q_I^2}{\alpha^2 f_Iq_I \prod_{J\neq I} (q_J-q_I)}\,,\qquad
\widetilde X = 1 - \fft{\alpha^2}{g^2} X\,.\label{stuboundary}
\ee
Even though the parameter space is considerably more complicated than for the case of Einstein-Maxwell theory, the general structure is nevertheless the same.  The space is bounded by two adjacent roots of the function $X$, which we call $x_1$ and $x_2$ taking $x_1<x_2$, within which $X$ is non-negative and has a maximum $X^{\rm max}$.  Thus, for sufficiently large $g$, $\widetilde X$ can be positive definite.  The solution then describes a bounded spacetime with a naked conical singularity at the south pole $x=x_1$, while a potential conical singularity at the north pole can be avoided by choosing the appropriate period for the angular coordinate $\varphi$.  On the other hand, for small enough $g$, namely $g^2/\alpha^2 \le X^{\rm max}$, the function $\widetilde X$ can develop a simple root $x_0$, satisfying $x_1<x_0<x_2$. Then the conical singularity at the south pole is hidden by the event horizon at $x=x_0$. In the extremal case for which $x_0$ becomes a double root, the
near-horizon geometry is AdS$_2\times S^1$.

\subsection{Single-scalar two-vector theories}

For the C-metric solutions in the one-parameter family of theories discussed in section 4, the boundary metric at $y=x$ can be put in the same form as (\ref{stuboundary}) with $X$ given by (\ref{specifyX}) and
\be
{\cal F}=f_1^{-\fft12 N_1} f_2^{-\fft12 N_2}\,,\qquad \widetilde X={\cal F}^{-1}-\fft{\alpha^2}{g^2} X\,.
\ee
Once again, the general structure is the same as for the previous cases. The space is bounded by $x=-1$ and $x=1$ and there is a conical singularity at one of the poles, which is hidden by an event horizon that is present for small enough $g$. In the extremal case, the near-horizon geometry is AdS$_2\times S^1$. As before, for certain values of the parameters, the boundary metric can be that of a smooth soliton despite there being a naked power-law singularity in the bulk.

\subsection{Dynamical black holes}

For the dynamical C-metric solution in section 5.1, the boundary metric is given by
\be
ds_3^2 =-2\eta\, du dx -\fft{g^2}{\alpha^2} Y du^2 + \fft{dx^2}{X} + X d\varphi^2\,,\qquad Y=f-\fft{\alpha^2}{g^2} X\,,
\ee
where $X$ and $f$ are given in (\ref{time-dependent-solution}). This is dual to a strongly-coupled three-dimensional conformal field theory on the background of an evolving black hole.

Since the global structure of this evolving black hole is complicated, it is instructive to study the static limits at $\alpha=\alpha_{1,2}$. Then the metric takes the form
\be
ds_3^2=\fft{f}{XY}\ dx^2-\fft{g^2}{\alpha^2} Y dt^2+X d\varphi^2\,,\qquad dt=\eta du+\fft{\alpha^2}{g^2Y} dx\,.
\ee
The space is bounded by $x=-x_0\equiv -m_2^2/m_1^2$ and $x=1$ and there is a conical singularity at $x=-x_0$, with the exception of $x_0=\ft13$ in which case conical singularities can be avoided at both poles. For the $\alpha=\alpha_2$ static solution, there is also a power-law curvature singularity at $x=-x_0$, even though it also coincides with a Killing horizon. For sufficiently large $\alpha_2$, the singularity can be hidden inside a horizon at $x_1$ with $-x_0<x_1<1$. For the $\alpha=\alpha_1$ static solution, the solution describes a soliton with $x\in [-x_0,1]$ with a conical singularity at $x=-x_0$, except for when $x_0=\fft13$.  For large enough $\alpha_1$, there is a horizon at $x_2$ with $x_2<x_1$. Thus, depending on the parameters, the boundary metric evolves from a black hole to either another black hole or else a (smooth) soliton. The singularities, regardless of whether they are of the conical or power-law type, can be hidden inside a horizon for an appropriate choice of parameters.

\section{C-metrics in five dimensions}

Generalizing C-metrics to higher dimensions turns out to be difficult.
A naive attempt is to take the ansatz
\be
ds^2=\fft1{\alpha^2 (x-y)^2}\Big[Y d\Omega_{p}^2 + \fft{dy^2}{Y} +
\fft{dx^2}{X} + X d\Omega_q^2\Big]\,,
\ee
where $d\Omega_{p}^2$ and $d\Omega_q^2$ are the metrics for round spheres in $p$ and $q$ dimensions, respectively. Note that we are considering Euclidean signature here.  As long as one of the $(p,q)$ is bigger than 1, the equations of motion $R_{\mu\nu}+ (D-1)g^2 g_{\mu\nu}=0$ imply that
\be
X=\Xi(x)\,,\qquad Y=-\Xi(y) + \fft{g^2}{\alpha^2}\,,\qquad {\rm with} \qquad\Xi(\xi) = b_0 + 2b_1 \xi + b_2 \xi^2\,,
\ee
where the parameters satisfy $b_0 b_2-b_1^2 +1=0$. However, these metrics are maximally-symmetric, i.e.~they are locally either (A)dS or flat.  The global structure of these metrics with $p=1$ in the $(x,y)$ coordinates were analysed in \cite{Meng:2012zt}.

We will now construct a five-dimensional Ricci-flat metric by lifting the four-dimensional dilaton C-metric using the Kaluza-Klein reduction ansatz. The charged C-metrics we obtained in section 4 allow us to set the electric charges and the $g$ parameter to zero independently.  The result is a C-metric in Einstein gravity with a free scalar, namely
\be
e^{-1}{\cal L}_4 = R - \ft12 (\partial\varphi)^2\,.\label{einfree}
\ee
The solution is given by
\bea
ds_4^2 &=& \fft{1}{\alpha^2(x-y)^2} \Big( x^{\fft{N_1}2} (-Y d\tau^2 + \fft{dy^2}{Y}) + y^{\fft{N_1}{2}} (\fft{dx^2}{X} + X d\varphi^2)\Big)\,,\nn\\[1mm]
e^\phi &=& \Big(\fft{y}{x}\Big)^{\fft12\sqrt{N_1N_2}}\,,\qquad N_1+N_2=4\,,\nn\\[1mm]
X&=&(b_0 + b_1 x + b_2 x^2) x^{1-\fft12 N_1}\,,\qquad
Y=-(b_0 + b_1 y + b_2 y^2) y^{1-\fft12 N_1}\,.
\eea
Note that, in the above presentation, we set $q_2=0$, let $f_1\rightarrow x$ and $h_1\rightarrow y$ by making constant shifts of the coordinates $(x,y)$, and made further reparametrizations and coordinate rescalings. The theory (\ref{einfree}) can be lifted to five dimensions to become pure Einstein gravity.  The corresponding $D=5$ Ricci-flat metric is then given by
\bea
ds_5^2 &=& \fft{1}{(x-y)^2} \Big(\fft{y}{x}\Big)^{\fft12 \sqrt{\fft{N_1N_2}3}} ds_4^2 + \Big(\fft{x}{y}\Big)^{\sqrt{\fft{N_1N_2}3}} d\psi^2\,,\nn\\[1mm]
ds_4 &=& x^{\fft{N_1}2} y^{\fft{N_1-2}2}\Big(-y^{2-N_1}Y d\tau^2 + \fft{dy^2}{Y}\Big) + x^{\fft{N_1-2}2} y^{\fft{N_1}{2}} \Big(\fft{dx^2}{X} + x^{2-N_1} X d\phi^2\Big)\,,\nn\\[1mm]
X&=& c(x-a)(x-b)\,,\qquad Y=-c(y-a)(y-b)\,.\label{d5genN1}
\eea
In the above, we have stripped off the trivial parameter $\alpha$.  Furthermore, $(X,Y)$ are not the same as those defined in the earlier solutions.  Note that the metric is invariant under $x\rightarrow 1/y$, $y\rightarrow 1/x$ and $N_1\leftrightarrow N_2$, together with interchanging the role of $\tau$ and $\phi$.

Let us consider the simple parameter choice $N_1=1$ and hence $N_2=3$, for which there is no irrational power in $(x,y)$ coordinates. Then the solution (\ref{d5genN1}) can be written as
\bea
ds_5^2 &=& \fft{1}{(x-y)^2}
 \Big(-y  Y d\tau^2 + \fft{dy^2}{ Y} + \fft{y}{x} (\fft{dx^2}{X} + x X d\phi^2)\Big)
 + \fft{x}{y}\, d\psi^2\,,\nn\\[1mm]
X&=&c(x-a)(x-b)\,,\qquad Y=-c(y-a)(y-b)\,.
\eea
This metric can be obtained in a certain limit of the five-dimensional black ring metric \cite{Emparan:2001wn} or the more general class of metrics obtained in \cite{Lu:2008js}

A few of the curvature polynomial invariants are given below:
\bea
I_2 &\equiv& R_{\mu\nu\rho\sigma}R^{\mu\nu\rho\sigma}\nn\\[1mm]
&=&\ft32 c^2 (1-\ft{x}{y})^4\Big(3 a^2 b^2 - 4 (a + b) x y^2 + 2 a b y (-2 x + y) + y^2 (8 x^2 - 4 x y + 3 y^2)\Big)\,,\nn\\[1mm]
I_3 &\equiv& R^{\mu}{}_\nu{}^\rho{}_\sigma R^\nu{}_\mu{}^{\alpha}{}_\beta R^\beta{}_\alpha{}^\sigma{}_\rho\nn\\[1mm]
&=&3c^3 (1-\ft{x}y)^6 (-ab +y(2x-y))\Big(a^2b^2 -2(a + b) x y^2 +y^2(2x^2 + y^2)\Big)\,,\nn\\[1mm]
J_3 &\equiv& R^{\mu}{}_\nu{}^\rho{}_\sigma R^\alpha {}_\mu{}^{\beta}{}_\rho R^\nu{}_\alpha{}^\sigma{}_\beta
= -\ft12I_3\,.
\eea
This demonstrates that the metric is indeed of cohomogeneity two. The curvature is singular at $y=0$, $y=\infty$ and $x=\infty$ but not at $x=0$.  The curvature vanishes at $x=y$ except for $x=y=0$ and $\infty$. The metric acquires a coordinate singularity at the following locations: $x=0$ for which $g_{\psi\psi}=0$, $x=a$ and $b$ for which $g_{\phi\phi}=0$, and $y=a$ and $b$ for which $g_{\tau\tau}=0$.  Thus, a natural choice for the ranges on coordinates and parameters is
\be
x\in [0,a]\,,\qquad y\in [a,b]\,,\qquad b>a>0\qquad \hbox{and}\qquad c>0\,,
\ee
such that $X$ and $Y$ are non-negative and the $y$ coordinate lies between the two horizons at $y=a$ and $y=b$.

The asymptotic region is located at $x=a=y$. To analyze this region, it is useful to make the following coordinate transformation:
\be
\fft{\sqrt{a-x}}{y-x} =\ft12 \sqrt{c(b-a)}\, r \cos\theta\,,\qquad
\fft{\sqrt{y-a}}{y-x} = \ft12 \sqrt{c(b-a)}\, r\sin\theta\,,
\ee
and then take the limit $r\rightarrow \infty$.  In order for the metric to be properly normalized in the asymptotic region without rescaling the $\tau$ and $\phi$ coordinates, we find that
\be
c=\fft{2}{(b-a)\sqrt{a}}\,.
\ee
The asymptotic metric then takes the form
\be
ds_5^2 = dr^2 + r^2 (d\theta^2 + \cos^2 \theta d\phi^2 - \sin^2\phi d\tau^2) + d\psi^2\,.
\ee
This is Mink$_4\times S^1$ where the Mink$_4$ is written in a Rindler-like accelerating frame.

Conical singularities can be avoided at both $x=a$ and $b$ if the $\phi$ and $\psi$ coordinates have the periods
 \be
\Delta\phi=2\pi\,,\qquad \Delta\psi = \fft{4\pi}{\sqrt{c a b}}\,.
\ee
The temperatures associated with the horizons at $y=a$ and $b$ are given by
\be
T_{a}=\fft{1}{2\pi}\,,\qquad T_{b}=\sqrt{\ft{b}{a}}\,T_{a}> T_{a}\,.
\ee
The horizon at $y=a$ has infinite area, which is consistent with it being the accelerating horizon, and hence $T_a$ is the Unruh temperature.  The event horizon at $y=b$ has the topology of a distorted 3-sphere, with the metric
\be
(ds^2)_{\rm horizon}^{y=b} = \fft{b}{x(b-x)^2} \left(\fft{dx^2}{X} + x X d\phi^2\right) + \fft{x}{b} d\psi^2\,.
\ee
This horizon has finite area, giving rise to the Bekenstein-Hawking entropy
\be
S_b=\fft{2\pi^2 \sqrt{a}}{(b-a)b\sqrt{c}}\,.
\ee

As previously discussed, all of the four-dimensional C-metrics suffer from having naked conical singularities.  What is interesting is that our five-dimensional Ricci-flat metric may have no conical singularity at all. In addition, the power-law singularities at $y=0$ and $y=\infty$ are both hidden behind the horizons. Furthermore, the radius of the fifth coordinate $\psi$ remains finite in the region between the horizons.  It is more natural to define
\be
\tilde\psi=\ft12 \sqrt{cab}\, \psi\,,
\ee
so that $\Delta\tilde\psi=2\pi$.  In between the horizons, the maximum radius for $\tilde\psi$ occurs in the asymptotic region $y=a=x$ and is given by
\be
R^{\rm max}_{\tilde \psi}=\fft{2}{\sqrt{cab}}\,.
\ee
Thus, the coordinate $\psi$ is truly a compactifying coordinate.

It is worth remarking that, with an increasing value of $b$, the temperature $T_b$ increases whilst the entropy $S_b$ decreases. In the limit $b\rightarrow \infty$,
%we have
%\be
%X=\fft{2}{\sqrt{a}} (a-x)\,,\qquad Y=\fft{2}{\sqrt{a}} (y-a)\,,\label{schw}
%\ee
%and
the metric turns out to be locally the Schwarzschild black hole.  To see this, we define $a=4/\mu^2$ and make the coordinate transformation\footnote{We are grateful to Chris Pope for this coordinate transformation.}
\be
x=\fft{4(r^2-\mu)}{\mu^2 (r^2 - \mu\sin^2\theta)}\,,\qquad
y=\fft{4r^2}{\mu^2 (r^2 - \mu\sin^2\theta)}\,,
\ee
so that the metric can be written as
\be\label{schw}
ds_5^2 = \fft{dr^2}{f} +r^2 (d\theta^2 + \cos^2\theta\, d\phi^2 - \sin^2\theta\, d\tau^2) + f\,d\psi^2\,,\qquad f=1 - \fft{\mu}{r^2}\,.
\ee
The solution becomes the Euclidean Schwarzschild soliton if we Wick rotate the $\tau$ coordinate.  In the $(x,y)$ coordinates, the Euclidean soliton lives in the region specified by $x\in [0,a]$ and $y\in [a,\infty)$.

Alternatively, we can perform a double Wick rotation such that $\psi$ becomes timelike and $\tau$ becomes a spacelike compact coordinate, as in the case of the black ring \cite{Emparan:2001wn}.  Then the metric (\ref{schw}) is precisely that of the Schwarzschild black hole.  However, for the general cohomogeneity-two solution with $(a,b)$ parameters, one of the conical singularities associated with the collapsing of the now spatial direction $\tau$ at $y=a$ and $b$ cannot be avoided.  For the analogous reason, we shall not get smooth Euclidean metrics without conical singularities.

We end this section with a comment on the metric (\ref{d5genN1}) with general $N_1$, for which there is a curvature singularity at $x=0$, unlike the case of $N_1=1$. This can be avoided by choosing $x\in [a,b]$, with the penalty that the conical singularities at $x=a$ and $b$ cannot both be removed. Furthermore, at least one of the power-law singularities at $y=0$ and $y=\infty$ is naked.  While the conical singularities at $x=a$ and $b$ can be avoided for $N_1=2$, one of the power-law singularities at $y=0$ and $y=\infty$ remains naked.

\section{Conclusions}

We have constructed various generalizations of the C-metric, which describe black holes accelerating apart due to the presence of a cosmic string. This includes charged C-metric solutions in the STU gauged supergravity model that involve four independent electric charges, three (nontrivial) scalar charges and an acceleration parameter. In contract with previously-known C-metric solutions, conical singularities can be completely absent for a range of parameters, rather than for only a single point in parameter space. In certain limits, these solutions reduce to electrically neutral C-metrics supported solely by the scalar charges, as well as charged AdS black holes. Two-charge C-metric solutions have also been presented for a one-parameter family of Einstein-Maxwell-dilaton theories whose scalar potential can be expressed in terms of a superpotential.

In addition, we have discussed the dynamical generalizations of the dilaton C-metric and dilaton Ernst solutions. The time evolution is driven by the dilaton field, which transfers energy to the black holes. This causes the masses of the black holes to increase and their acceleration to decrease. We find that there are solutions which describe a pair of accelerating black holes evolving to a single accelerating black hole. For an appropriate value of the magnetic field in the case of the dynamical dilaton Ernst solution, the cosmic string can be completely removed for all times, such that the magnetic field performs the role of providing the necessary force to accelerate the black holes.

The metrics on the boundary of the C-metric solutions describe three-dimensional static black holes, gravitational solitons, or else a black hole evolving into either another black hole or a soliton. For an appropriate choice of parameters, all conical and power-law curvature singularities are hidden inside a horizon. Interestingly enough, in certain cases the bulk spacetime can have a naked power-law singularity while there is nevertheless a smooth soliton on the boundary. Applying the AdS/CFT correspondence to these backgrounds yields a holographic description of a strongly-coupled three-dimensional conformal field theory on the background of a static black hole, a gravitational soliton, or an evolving black hole. An analysis of the dynamical black funnels and droplets contained within the bulk solutions would provide insight as to whether the gravitational backgrounds on the boundary are strongly or weakly coupled to the field theory \cite{Hubeny:2009ru,Hubeny:2009kz}.

For rather restricted cases, the solutions can be lifted to Ricci-flat metrics in five dimensions. It would be interesting to generalize the resulting five-dimensional solutions away from the pure gravitational sector, as well as to find higher-dimensional analogs of the C-metric.

\bigskip
\noindent{\bf Acknowledgements}:

\noindent We are grateful to Chris Pope and Zhibai Zhang for useful discussions. The research of H.L. is supported in part by NSFC grants 11175269, 11475024 and 11235003.

\end{document}